\documentstyle[twocolumn,seceq,epsf]{jpsj}

\def\runtitle{Equilibrium Chemical Engines}
\def\runauthor{Tatsuo {\sc Shibata} and Shin-ich {\sc Sasa}}

\title{Equilibrium Chemical Engines}

\author{
Tatsuo {\sc Shibata}\footnote{shibata@complex.c.u-tokyo.ac.jp} and
Shin-ichi {\sc Sasa}\footnote{sasa@jiro.c.u-tokyo.ac.jp}}
\inst{
Department of Pure and Applied Sciences, University of Tokyo.\\
Komaba, Meguro-ku, Tokyo 153, Japan}

\recdate{\today; version 1.0}

\abst{
An equilibrium reversible cycle with a certain engine to transduce the
energy of any chemical reaction into mechanical energy is proposed.
The efficiency for chemical energy transduction is also defined so as
to be compared with Carnot efficiency.  Relevance to the study of
protein motors is discussed.}

\kword{Chemical thermodynamics, Engine, Efficiency, Molecular machine.}

\begin{document}
\sloppy
\maketitle

\section{Introduction}
\label{sec:Introduction}

The molecular machine in biological systems works as a chemical energy
transducer.  For instance~\cite{CELL}, the protein motors such as
myosin and kinesins, or flagellar motor produce mechanical energy as
they consume the chemical energy provided by hydrolysis of ATP or
proton density gradient.  Some other proteins, such as enzymes or pump
proteins of membrane transportation, can be also thought as chemical
energy transducer in which energy input and output are both chemical.
Recently various phenomenological models for protein motors have been
studied and the thermodynamic interpretation for these models have
been presented~\cite{Sekimoto97I,Sekimoto97II}.  However, the models,
in which chemical reaction is taken into account, are scarcely
proposed.  While the energetics of models, in which chemical reaction
is explicitly supposed, may be future task~\cite{Sasa}, none have
discussed an equilibrium cycle associated with any chemical reaction,
as Carnot considered Carnot cycle for thermal energy transduction.
Also the notion of energy transduction efficiency from chemical to
mechanical has never been stated explicitly. These will be important
when we compare the energy transduction in molecular machines with
macroscopic thermodynamics.

In this Paper, we discuss how we should define a proper chemical
energy transduction efficiency and how we construct an engine which
gives its maximal value.  In \S 2, a chemical engine to transduce
energy of arbitrary chemical reaction into mechanical energy is
proposed.  In \S 3, we review the chemical thermodynamics of the
processes involving the chemical reaction. Equilibrium chemical
reaction is also presented.  In \S 4 and \S 5, we show an equilibrium
reversible cycle using the chemical engine as to correspond to the
idealized cycle for heat engines which Carnot considered. Further we
propose efficiency for chemical energy transduction, and show its
maximum value.  In \S 6, the heat flow associating with particle flow
is studied.  In \S 7, we study some applications to compare energetic
behavior of molecular engines with macroscopic engines.  The paper
concludes in \S 8 with a discussion about efficiency.

\section{Chemical Engine}
\label{sec:Chemical Engine}

Consider $n$ chemical components among which chemical reaction
\begin{equation}
\sum_{i}^{n} \nu_{i} \mbox{M}_{i} \rightleftharpoons 0
\label{eq:chemreact}
\end{equation}
takes place, where $\mbox{M}_{i}$ is the symbol for the $i$'th
chemical component, and $\nu_{i}$ is stoichiometric coefficient which
is positive when the component is a product in the chemical reaction,
while negative when the component is a reactant. For convenience, let
the chemical formula be written as the spontaneous reaction takes
place in the direction that the products increase when the chemical
affinity
\begin{equation}
A=- \sum_{i} \nu_{i} \mu_{i}
\end{equation}
is positive, where $\mu_i$ is the chemical potential of the $i$'th
component.  

\begin{figure}
\epsfysize=.4\textwidth
\centerline{\epsffile{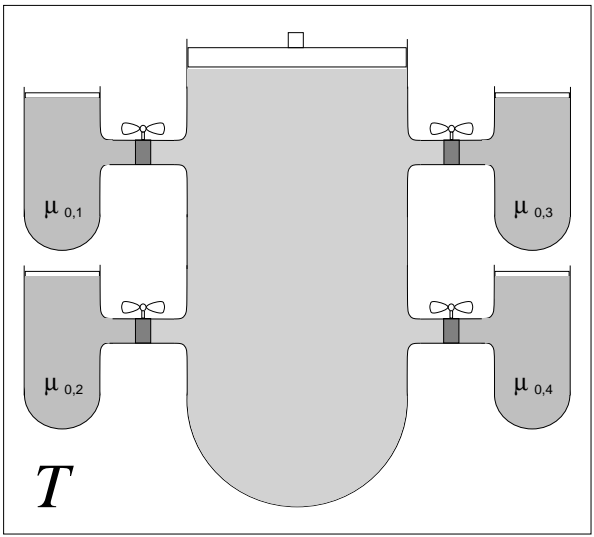}}
\caption{schematic figure of equilibrium chemical engine.
The center of this figure is a container which has a piston to
generate mechanical energy. The container also has sources in each of
which only one of the components associating with chemical reaction is
contained.}
\label{fig:ChemEngine}
\end{figure}

We enclose the components in a container(reaction box), which has a
piston to change the volume $V$ and the pressure $p$ in touch with a
heat bath, so that the chemical reaction takes place and the energy of
chemical reaction is transduced into mechanical energy (see
Fig.\ref{fig:ChemEngine}).

The container can also take any $i$'th component in and out from
source which contains only the $i$'th component at the chemical
potential $\mu_{0,i}$. We assume that an ideal semi permeable membrane
to the $i$'the component which separate the container and the source
is completely impermeable to all other components.  This configuration
is similar to Van't Hoff reaction box~\cite{Fermi}.

\section{Equilibrium Chemical Reaction}
\label{sec:Equilibrium Chemical Reaction}

We review the chemical thermodynamics of processes involving chemical
reaction eq.(\ref{eq:chemreact}).

The change of Gibss's free energy $G=G(T,p,\{N_{i}\})$ in the
container is given by
\begin{equation}
dG=-SdT+Vdp+\sum_{i}\mu_{i}dN_{i},
\label{eq:gibbs}
\end{equation}
where $N_{i}$ is the number of molecules of $i$'th component.

We first discuss the process without exchange of components with
sources.  The change of $N_{i}$ is given by
\begin{equation}
dN_{i}=\nu_{i}d\xi,
\label{eq:reaction}
\end{equation}
where $d\xi$ is parameter for extent of reaction.
Substituting eq.(\ref{eq:reaction}) into eq.(\ref{eq:gibbs}) yields
\begin{equation}
dG=-SdT+Vdp+\sum_{i}\mu_{i}\nu_{i}d\xi.
\label{eq:gibbs'}
\end{equation}
When $T$ and $P$ are changed quasi-statically, chemical reaction
occurs with shifting an equilibrium state continuously.  In such a
process, the chemical equilibrium condition
\begin{equation}
\sum_{i}\nu_i\mu_i=0
\label{eq:chemeq}
\end{equation}
holds.  Then eq.(\ref{eq:gibbs'}) becomes
\begin{equation}
dG=-SdT+Vdp.
\label{eq:gibbs1}
\end{equation}
Since $\mu_{i}=\mu_{i}(T,p,\xi)$, and the chemical equilibrium
condition eq.(\ref{eq:chemeq}), the extent of reaction is derived as a
function of $T$ and $p$,
\begin{equation}
\xi=\xi(T,p).
\end{equation}
We call such chemical reaction in a quasi-static process ``equilibrium
chemical reaction''.

Next we consider exchanging the $j$'th component with the source whose
chemical potential is $\mu_{0,j}=constant$.  When $dN_{j}^{ext}$
denotes the number of molecules of $j$'th component which the
container absorbs from the source, we can extend
eq.(\ref{eq:reaction}) to the form 
\begin{equation}
dN_{j}=\nu_{j}d\xi+dN_{j}^{ext}.
\label{eq:exreaction}
\end{equation}
Therefore from eqs.(\ref{eq:gibbs}) and (\ref{eq:chemeq}) we obtain
\begin{equation}
dG=-SdT+Vdp+\mu_{0,j}dN_{j}^{ext}.
\label{eq:gibbs2}
\end{equation}
By the two equilibrium conditions $\mu_{j}=\mu_{0,j}$ 
and eq.(\ref{eq:chemeq}),
the number of independent variables 
in equation $\mu_{i}=\mu_{i}(T,p,\xi,N_{j}^{ext})$
is reduced to two.
In this way,
we can regard $\xi$, and
$N_{j}^{ext}$ as dependent variables of $T$ and $p$.
We thus obtain
\begin{equation}
dG=\left(-S+\mu_{0,j} 
\frac{\partial N_j^{ext}}{\partial T}\right)dT
+\left(V+\mu_{0,j}\frac{\partial N_j^{ext}}{\partial p}\right)dp.
\end{equation}

\section{Equilibrium Reversible Cycle}
\label{sec:Equilibrium Reversible Cycle}

We show how an equilibrium reversible cycle $C$ is realized by
employing the chemical engine presented in \S 2. In the argument
below, we suppose that the container is always kept in touch with a
heat bath and that all processes are quasi-static.

At the initial state $\mu_{1}$ is assumed to equal to $\mu_{0,1}$.
While the equilibrium conditions is kept, the volume $V$ is shifted so
that the container can absorb the first component from the source.
Next the source of the $i$'th component is separated from the
container and the volume $V$ is changed until $\mu_2$ equals to
$\mu_{0,2}$.  After that, let the container touch with the source for
the second component, and the volume $V$ is shifted so that the
container can absorb the second component, as in the case of first
component, and so forth.

Repeating the similar procedure, we have the following cycle $C$,
which consists of $2n$ steps.  let $i$ be $1,2,\cdots, n$.
\begin{itemize}
\item \underline{$2i-1$'th step:}
Let the container be kept in touch with the source of the $i$'th component.
While the equilibrium conditions is kept,
the volume $V$ is changed so that the container can absorb the component.
Note that an amount of the component absorbed into the container is positive
when the component is reactant while negative when product.

\item \underline{$2i$'th step:}
Separating the source from the container, the volume $V$ is changed
until $\mu_{i+1}$ comes to be $\mu_{0,i+1}$.
\end{itemize}
The $i$ is assumed to set in order, but the order is allowed to set
arbitrary.  After $2n$'th step, one cycle will be completed by
changing $\mu_1$ until $\mu_1$ equals $\mu_{0,1}$, so as to bring the
system back to its initial state.

\section{Efficiency}
\label{sec:Efficiency}

We discuss energy transduction efficiency for chemical engines.  First
we notice that we should define the efficiency so as to compare
chemical engines with heat engines.  
In the case of heat engines
with two heat baths of different temperature ($T_{+}>T_{-}$), 
the efficiency for heat engines was defined as
\begin{equation}
\eta_{heat}=\frac{W}{Q_{+}},
\end{equation}
where the system absorbs the heat $Q_{+}$ from the heat bath and
transduces into the mechanical energy $W$
The Carnot cycle gives the maximum efficiency
\begin{equation}
\eta_{heat}=\frac{T_{+}-T_{-}}{T_{+}}.
\end{equation}

\begin{figure}[t]
\epsfysize=.22\textwidth
\centerline{\epsffile{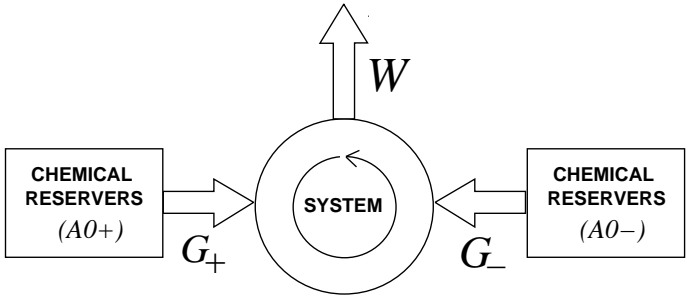}}
\caption{schematic figure of free energy flow.
The container absorbs the free energy
$G_{+}=\sum_{\nu_{i}>0}\mu_{i}N_{i}^{ext}$ from the sources of
products($\nu_{i}>0$) while from reactant($\nu_{i}<0$)
$G_{-}=\sum_{\nu_{i}<0}\mu_{i}N_{i}^{ext}$.  In the figure the sources
of reactants is denoted by $A_{0-}=\sum_{\nu_{i}<0}\nu_{i}\mu_{0,i}$,
while the sources of products is denoted by
$A_{0+}=-\sum_{\nu_{i}>0}\nu_{i}\mu_{0,i}$. }
\label{fig:FreeEnergyFlow}
\end{figure}

In the case of chemical engines,
the efficiency should be defined as 
\begin{equation}
\eta = \frac{W}{G_{+}}
\label{eq:eff}
\end{equation}
with the free energy $G_{+}=\sum_{\nu_{i}>0}\mu_{0,i}N_{i}^{ext}>0$
absorbed by the container from the sources of the
products($\nu_{i}>0$), where the container absorbs an amount
$N_{i}^{ext}$ of the $i$'th source (see Fig.\ref{fig:FreeEnergyFlow}).

We now calculate  the work $W$ and $G_{+}$ for the cycle $C$.
Since the cycle $C$ works isothermally, thermodynamic relation for
each steps is written as
\begin{itemize}
\item \underline{$2i-1$'th step}: 
\begin{equation}
dG = Vdp+\mu_{0,i}dN_{i}^{ext},
\label{eq:thermo1}
\end{equation}
\item \underline{$2i$'th step}: 
\begin{equation}
dG = Vdp.
\label{eq:thermo2}
\end{equation}
\end{itemize}
where we have used eqs.(\ref{eq:gibbs1}) and (\ref{eq:gibbs2}).

From eqs.(\ref{eq:thermo1}) and (\ref{eq:thermo2}) we get
\begin{equation}
\oint dG = 
\oint Vdp + \sum_{i}^{s}\int_{2i-1}^{2i}\mu_{0,i} dN_i^{ext}=0, 
\label{eq:cycle}
\end{equation}
where the integral is taken over one cycle. Thus the work $W$ in the
cycle $C$ is given by
\begin{equation}
W  =  - \oint Vdp\nonumber
   =  \sum_{i}^{n}\mu_{0,i}\int_{2i-1}^{2i}dN_i^{ext}.
\label{eq:work0}
\end{equation}

Since $N_{i}$ brings back to its initial value after one cycle, we
obtain
\begin{equation}
\oint dN_{i}=
\nu_i \oint d\xi + \int_{2i-1}^{2i}dN_i^{ext} = 0,
\label{eq:react}
\end{equation}
where we have used eq.(\ref{eq:exreaction}).

Substituting eq.(\ref{eq:react}) into eq.(\ref{eq:work0}) yields
\begin{equation}
W=-\sum_{i=1}^{n}\nu_i\mu_{0,i}\oint d\xi=A_{0}\oint d\xi,
\label{eq:work1}
\end{equation}
where $A_{0}$ is chemical affinity of sources
\begin{equation}
A_{0}=-\sum_{i}\nu_i\mu_{0,i}.
\end{equation}
Equation (\ref{eq:work1}) is the work within one cycle transformed
from the energy of chemical reaction.  Note that the result is
independent of ratio of components in the container, or nature of
working substance, e.g. it does not depend on it whether or not the
substance is ideal.

From eqs.(\ref{eq:react}) and (\ref{eq:work1}), 
the efficiency of the cycle $C$ is given by 
\begin{equation}
\eta  =  \frac{-\sum_{i}\nu_{i}\mu_{0,i}\oint d\xi}%
                {-\sum_{\nu_{i}>0}\nu_{i}\mu_{0,i}\oint d\xi} =
      \frac{A_{0+}-A_{0-}}{A_{0+}},
\label{eq:max}
\end{equation}
where $A_{0-}=\sum_{\nu_{i}<0}\nu_{i}\mu_{0,i}$, and
$A_{0+}=-\sum_{\nu_{i}>0}\nu_{i}\mu_{0,i}$.  Note that the chemical
affinity is decomposed as $A_{0}=A_{0+}-A_{0-}$.

If there were the cycle $C'$ whose efficiency is higher than
eq.(\ref{eq:max}), by coupling this cycle $C'$ with the above cycle
$C$, the chemical reaction could take place in a direction opposite to
that predicted by its own affinity, and as a result, the total free
energy in the sources increased spontaneously.  Since this violates
the 2nd thermodynamic law, the cycle $C$ gives the maximum efficiency.

\section{Heat Flow}
\label{sec:Heat Flow}

So far, we have been attention to the free energy flow
between the container and the sources in the cycle $C$.
We next consider the heat flow
among the container, the sources, and the heat bath.

\begin{figure}[t]
\epsfysize=.35\textwidth
\centerline{\epsffile{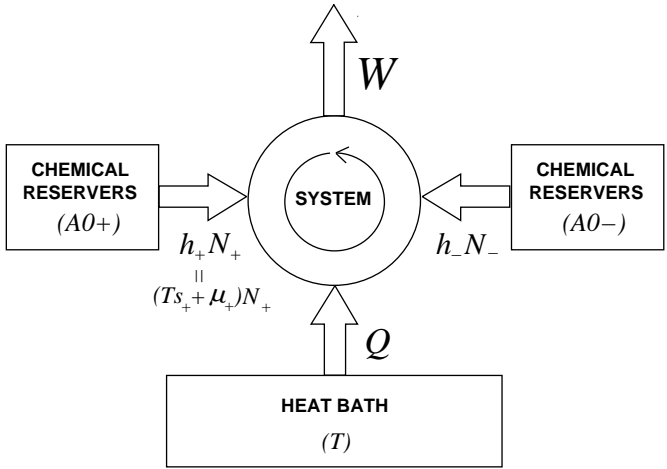}}
\caption{schematic figure of the heat flow.
Particle transportation $\mu N$ is inevitably accompanied by heat flow $TsN$.
The total energy absorbed by the container from the source 
equals to enthalpy $hN$.
With the heat $TsN$ from the source,
heat $Q$ is transported from the heat bath into the container.}
\label{fig:EnergyFlow}
\end{figure}

Processes in the sources and the heat bath are assumed to be quasi-static.
When the entropy per one molecule of the $i$'th component 
in the source is denoted by $s_{i}$,
the thermodynamic relation for the $i$'th source is given by
\begin{equation}
dU_{i}=-\left(Ts_{i}+\mu_{i}\right)dN_{i}^{ext},
\end{equation}
in which we suppose that changes of volume in the sources is so small that 
these are negligible.
We can include the effect of volume change if necessary.
In this formula, $(Ts_{i}+\mu_{i})$ can be interpreted as 
the enthalpy per one molecule of the $i$'th sources $h_{i}$.
The thermodynamic relation for the heat bath is given by
\begin{equation}
dU_{H}=TdS_{H}.
\end{equation}
Since the energy conservation law gives 
$\Delta U + dU_{i} + dU_{H} = W $
where $\Delta U$ denotes the energy change in the container,
we obtain
\begin{equation}
\Delta U=-TdS_{H}+h_{i}dN_{i}^{ext}-W.
\end{equation}
Thus, in one cycle,
\begin{equation}
W=Q+\sum_{i}h_{i}N_{i}^{ext},
\end{equation}
where $Q=-\oint TdS_{H}$ is heat flow from the heat bath.  We should
note that particle transportation is inevitably accompanied by heat
flow from the source (see Fig.\ref{fig:EnergyFlow}).

Under the equilibrium condition of the container, when the container
touches with the $i$'th source and the heat bath, the entropy change
in the heat bath is given by
\begin{equation}
dS_{H}=-dS+s_{i}dN_{i}^{ext}.
\label{eq:entropyheatbath}
\end{equation}
Note that the heat flow from the heat bath into the container is not $TdS$.
K. Kitahara has developed similar argument 
in the context of heat engine with particle flow~\cite{Kitahara}.
By integrating eq.(\ref{eq:entropyheatbath}), 
the heat absorbed by the container 
from the heat bath in one cycle $Q$ is written as
\begin{equation}
Q=-T\sum_{i}s_{i}N_{i}^{ext},
\label{eq:heat}
\end{equation}
which can be positive or negative depending on the detail of the
chemical reaction.  Note that because of the isothermal condition and
$\oint dS=0$, the total heat absorbed by the container is equal to
zero, whether the chemical reaction eq.(\ref{eq:chemreact}) is
exothermic or endothermic.  Therefore, the sources absorb the heat $Q$
from the container as expressed in the form of the right handed side
of eq.(\ref{eq:heat}).

\section{Application}
\label{sec:Application}

To analyze the difference between molecular
motors and macroscopic engines, molecular motors should be compared
with this idealized cycle, just as Feynmann's ratchet~\cite{Feynmann}
is compared with Carnot cycle.
The cycle as we have discussed above is the most ideal than any other
engines with chemical reaction including molecular motors such as
myosin or kinesins.  
For molecular motors with the hydrolysis reaction of
ATP$\rightarrow$ATP$+$P, the maximum efficiency becomes
\begin{equation}
\eta=\frac{\mu_{ATP}-\mu_{ADP}-\mu_{P}}%
               {-\mu_{ADP}-\mu_{P}}.
\end{equation}
Because chemical potential is negative and is increased with
concentration, high concentration of ATP in the living systems makes
this efficiency higher.

On the other hand, flagellar motors seems to consume the potential
energy in the proton density gradient without chemical reaction.  The
above result can be applicable to this case.
Suppose that the working substance consists of a simple chemical
component M and there are two sources whose chemical potentials
are $\mu_{-}$ and $\mu_{+}$.  
Even in such a case, we can regard particle transportation between two sources
as a sort of chemical reaction in the following way.
When a molecule in each source (with the values of $\mu_{-}$ and $\mu_{+}$)
is denoted by $M_{-}$  and $M_{+}$ respectively,
transportation between two sources can be expressed by
\begin{equation}
\mbox{M}_{-}\rightarrow\mbox{M}_{+}
\label{eq:SimpReact}
\end{equation}
in cycle, the maximum efficiency $\eta$ is
\begin{equation}
\eta=\frac{\mu_{+}-\mu_{-}}{\mu_{+}}.
\end{equation}

For molecular engines such as enzymes, pump proteins of membrane transport,
and mitochondrial electron transport chain,
several chemical reactions take place simultaneously.
We then consider a cycle in which several chemical reaction
take place simultaneously in the container.
Consider the $j$'th chemical reaction
\begin{equation}
\sum_{i=1}^{n_{j}} \nu_{ji} \mbox{M}_{ji} \rightleftharpoons 0,\,\,\,
(j=1, 2, 3, \cdots m),
\label{eq:chems.eq}
\end{equation}
where $\mbox{M}_{ji}$ is the $i$'th component of the $j$'th reaction, 
$\nu_{ji}$ is stoichiometric coefficient, 
and $m$ is the total number of reactions.
Then the work in one cycle is given by
\begin{equation}
W=\sum_{j=1}^{m}A_{0j}\oint d\xi_{j},
\end{equation}
with the chemical affinity of the $j$'th reaction
$A_{0j}=-\sum_{i=1}^{n_{j}}\nu_{ji}\mu_{0,ji} > 0$, where $\mu_{0,ji}$
is chemical potential of the source of $\mbox{M}_{ji}$, and $\oint
d\xi_{j}$ is extent of the reaction.  The maximum efficiency $\eta$ is
given by
\begin{equation}
\eta=\frac{\sum_{j=1}^{m}\left(A_{0j+}-A_{0j-}\right)\oint d\xi_{j}}%
{\sum_{j=1}^{m}A_{0j+}\oint d\xi_{j}},
\end{equation}
where $A_{0j-}=\sum_{\nu_{ji}<0}\nu_{ji}\mu_{0,ji}$, and
$A_{0j+}=-\sum_{\nu_{ji}>0}\nu_{ji}\mu_{0,ji}$.  Note that in this
case the maximum efficiency depends on how to operate the cycle.

With regards to the molecular engines with more than one chemical
reaction such as enzyme, pump proteins of membrane transport, and
mitochondrial electron transport chain, these might be thought as
chemical energy transducers which produce chemical energy rather than
mechanical energy.  When the container absorbs free energy $G_{j}$
from the sources for the components in the $j$'th chemical reaction,
$G_{j}$ is expressed by
\begin{equation}
G_{j}=\sum_{i}\mu_{ji}N_{ji}^{ext},
\end{equation}
where the container absorbs an amount $N_{ji}^{ext}$ of M$_{ji}$.
Therefore the inequality
\begin{equation}
\sum_{j=1}^{m}G_{j}\geq0
\end{equation}
should be satisfied, when the process is spontaneous.  Under this
condition, $G_{j}$ can be negative, and as a result the total free
energy in the sources associated with this $j$'th reaction increases
in one cycle.  Let us call the chemical reaction which satisfies
$G_{j}>0$ ``input chemical reaction'', while ``output chemical
reaction'' for $G_{j}<0$.  In this cycle, the energy of input chemical
reaction is transformed into the energy of output chemical reaction.
In other words, the container catalyzes the output chemical reaction
which can not take place spontaneously, utilizing the energy of input
chemical reaction.  The catalytic energy efficiency $\zeta$ may be
defined by
\begin{equation}
\zeta=\frac{-\sum_{G_{j}\leq0}G_{j}}{\sum_{G_{j}>0}G_{j}}.
\end{equation}
In quasi-static process, all the chemical energy of input chemical
reaction can be transformed into the energy of output chemical
reaction.  Thus the maximum catalytic efficiency is 1.  Note that the
stoichiometric relation between the components associated with the
different chemical reactions in general depends on the process.  For
instance, in the case of $m=2$, the ratio of the extent in the two
reaction $\oint d\xi_{1}\over\oint d\xi_{2}$ depends on process.

\section{Discussion}
\label{sec:Discussion}

Before concluding this Paper, we discuss the definition of efficiency.
The efficiency
\begin{equation}
\eta' = \frac{W}{G_{+}+G_{-}}
\label{eq:eff'}
\end{equation}
has been used in some bibliography~\cite{CELL}.  The maximum value of
this quantity turns to be 1 from the discussion in \S
\ref{sec:Efficiency}.  We should consider that this efficiency
eq.(\ref{eq:eff'}) can be used to compare the energy loss by
dissipation accompanied with irreversible heat generation in different
cycles under the {\it same} condition.  On the other hand, we consider
that the efficiency eq.(\ref{eq:eff}) presented in this Paper can be
applicable to compare different cycles under {\it different}
conditions, while the efficiency eq.(\ref{eq:eff}) can be also used
for cycles under the {\it same} condition.  Therefore, we believe the
latter definition eq.(\ref{eq:eff}) is preferable than former
eq.(\ref{eq:eff'}) to compare the efficiency of cycles.

Thermal ratchet engines such as Feynmann's ratchet~\cite{Feynmann} are
microscopic heat engines, which rectify thermal fluctuation to
generate mechanical energy.  Thermal fluctuation is believed to play
an important part in function of proteins. Thus the thermal ratchet
engines might bring some notions of energy transduction in
proteins~\cite{Vale}.  However, while high efficiency in molecular
motors pointed out, it has been shown that the efficiency of the
thermal ratchet engines is much less than Carnot
efficiency~\cite{Sekimoto97I}. While thermal energy transduction has
not been understood well, energy transduction with chemical reaction
and heat might be far beyond our comprehension.  In order to develop
notion of chemical energy transduction in molecular systems, it is a
future problem to construct the comprehensive model for motor proteins
in which chemical reaction should be explicitly supposed~\cite{Sasa}.
The efficiency we have proposed in this paper may be one of the useful
property to compare models with actual motor proteins.  We believe
that our study here may provide systematic tool for discussing the
efficiency in these systems.

\acknowledgements
One of the authors(T.S.) would like to thank S. Tasaki for his
stimulating talk at Hayama in 1995, and also grateful to K. Niizeki
and T. Hondo for some fruitful discussions and valuable comments at
Tohoku University.  The authors also gratefully acknowledge
K. Kitahara for sending his unpublished note and K. Kaneko for many
discussions and encouragement.  This work was partly supported by
grants from the Ministry of Education, Science, Sports and Culture of
Japan, No. 09740305 and from National Science Foundation,
No. NSF-DMR-93-14938.

\end{document}